\documentclass[conference]{IEEETran}
\IEEEoverridecommandlockouts
\usepackage{amssymb, amsmath, amsthm }
\usepackage{graphicx,subfig}
\usepackage{setspace, bbm}

\newcommand{\ba}{\begin{align*}}
\newcommand{\eaa}{\end{align*}}
\newcommand {\nt} {\notag}
\newcommand{\nl}{\notag\\}

\newcommand{\frn}{\frac 1 n}

\newcommand{\calX}{{\cal X}}
\newcommand{\hcalX}{{\hat{\cal X}}}
\newcommand{\calY}{{\cal Y}}
\newcommand{\calZ}{{\cal Z}}

\newcommand{\calA}{{\cal A}}

\newcommand{\calW}{{\cal W}}

\newcommand{\calR}{{\cal R}}

\newcommand {\bZ} {\mbox{\boldmath $Z$}}
\newcommand {\bu} {\mbox{\boldmath $u$}}

\newcommand {\bE} {\mbox{\boldmath $E$}}

\newcommand {\hX}{\hat{X}}
\newcommand {\tX}{\tilde{X}}

\newcommand {\hx}{\hat{x}}

\newcommand {\hZ}{\hat{Z}}

\newcommand{\eqd}{\stackrel{\triangle}{=}}

\newcommand{\sbr}[1] {\left[#1\right]}
\newcommand{\cbr}[1] {\left\{#1\right\}}

\newtheorem{theorem}{Theorem}

\begin{document}
\title{On Real--Time and Causal Secure Source Coding}
\author{\IEEEauthorblockN{Yonatan Kaspi and Neri Merhav$^\dag$}\thanks{$^\dag$This research
was supported by the Israeli Science Foundation (ISF) grant no.\ 208/08.}
\IEEEauthorblockA{
Department of Electrical Engineering \\
Technion - Israel Institute of Technology \\
Technion City, Haifa 32000, Israel\\
Email: \{kaspi@tx, merhav@ee\}.technion.ac.il}}


\maketitle

\begin{abstract}
We investigate two source coding problems with secrecy constraints. In the first problem we consider real--time fully secure transmission of a memoryless source. We show that although classical variable--rate coding is not an option since the lengths of the codewords leak information on the source, the key rate can be as low as the average Huffman codeword length of the source. In the second problem we consider causal source coding with a fidelity criterion and side information at the decoder and the eavesdropper. We show that when the eavesdropper has degraded side information, it is optimal to first use a causal rate distortion code and then encrypt its output with a key.
\end{abstract}

\section{Introduction}
We consider two source coding scenarios in which an encoder, referred to as Alice, transmits outcomes of a memoryless source to a decoder, referred to as Bob. The comunnication between Alice and Bob is intercepted by an eavesdropper, referred to as Eve.

In the first scenario, we consider real--time communication between Alice and Bob and require full secrecy, meaning that the intercepted transmission does not leak any information about the source. In the second scenario, we consider lossy causal source coding when both Bob and Eve have access to side information (SI). We require that Eve's uncertainty about the source given the intercepted signal and SI will be higher than a certain threshold.

Real--time codes are a subclass of causal codes, as defined by Neuhoff and Gilbert \cite{NeuhoffGilbert1982}. In \cite{NeuhoffGilbert1982}, entropy coding is used on the whole sequence of reproduction symbols, introducing arbitrarily long delays. In the real--time case, entropy coding has to be instantaneous, symbol--by--symbol (possibly taking into account past transmitted symbols). It was shown in \cite{NeuhoffGilbert1982}, that for a discrete memoryless source (DMS), the optimal causal encoder consists of time--sharing between no more than two memoryless encoders. Weissman and Merhav \cite{TsachyNeri2005} extended \cite{NeuhoffGilbert1982} by including SI at the decoder, encoder or both.

Shannon \cite{Shannon1949} introduced the information-theoretic notion of secrecy, where security is measured through the remaining uncertainty about the message at the eavesdropper. This information-theoretic approach of secrecy allows to consider security issues at the physical layer, and ensures unconditionally (regardless of the eavesdroppers computing power and time) secure schemes, since it only relies on the statistical properties of the system. Wyner introduced the wiretap channel in \cite{Wyner1975} and showed that it is possible to send information at a positive rate with perfect secrecy as long as Eve's channel is a degraded version of the channel to Bob. When the channels are clean, two approaches can be found in the literature of secure communication. The first assumes that both Alice and Bob agree on a secret key prior to the transmission of the source (through a separate secure channel for example). The second approach assumes that Bob and Eve (and possibly Alice) have different versions of side information and secrecy is achieved through this difference.

For the case of shared secret key, Shannon showed that in order for the transmission of a DMS to be fully secure, the rate of the key must be at least as large as the entropy of the source. Yamamoto (\cite{Yamamoto1997} and references therein) studied various secure source coding scenarios that include extension of Shannon's result to combine secrecy with rate--distortion theory. In both \cite{Shannon1949},\cite{Yamamoto1997}, when no SI is available, it was shown that separation is optimal. Namely, using a source code followed by encryption with the shared key is optimal. The other approach was treated more recently by Prabhakaran and Ramchandran \cite{Prab-Ramch1997} who considered lossless source coding with SI at both Bob and Eve when there is no rate constraint between Alice and Bob. It was shown that the Slepian-Wolf \cite{SlepianWolf73} scheme is not necessarily optimal when the SI structure is not degraded. Coded SI at Bob and SI at Alice where considered in \cite{GunduzEkripPoor2008}. These works were extended by Villard and Piantanida \cite{VillardPaint2011} to the case where distortion is allowed and coded SI is available to Bob. Merhav combined the two approaches with the wire--tap channel \cite{Merhav2008}. Note that we mentioned only a small sample of the vast literature on this subject.

In the works mentioned above, there were no constraints on the delay and/or causality of the system. As a result, the coding theorems of the above works introduced arbitrary long delay and exponential complexity.

The practical need for fast and efficient encryption algorithms for military and commercial applications along with theoretical advances of the cryptology community, led to the development of efficient encryption algorithms and standards which rely on relatively short keys. However, the security of these algorithms depend on computational complexity and the intractability assumption of some hard problems. To the best of our knowledge, there was no attempt so far to analyze the performance of a real--time or causal secrecy system from an information theoretic point of view.

The extension of Neuhoff and Gilbert's  result \cite{NeuhoffGilbert1982} to the real--time case is straightforward and is done by replacing the block entropy coding by instantaneous Huffman coding. The resulting bitstream between the encoder and decoder is composed of the Huffman codewords. However, this cannot be done when secrecy is involved, even if only lossless compression is considered. To see why, consider the case where Eve intercepts a Huffman codeword and further assume the bits of the codeword are encrypted with a one--time pad. While the intercepted bits give no information on the encoded symbol (since they are independent of it after the encryption), the number of intercepted bits leaks information on the source symbol. For example, if the codeword is short, Eve knows that the encrypted symbol is one with a high probability (remember that Eve knows the source statistics). This suggests that in order to achieve full security, the lengths of the codewords emitted by the encoder should be independent of the source.

In the last example, we assumed that Eve is informed on how to parse the bitstream into separate codewords. This will be the case, for example, when each codeword is transmitted as a packet over a network and the packets are intercepted by Eve. Even if the bits are meaningless to Eve, she still knows the number of bits in each packet. We show in the sequel that, albeit the above example, the key rate can be as low as the average Huffman codeword length (referred hereafter as the Huffman length) of the source. Full secrecy, in this case, will be achieved by randomization at the encoder, which can be removed by Bob. In contrast to the works mentioned above, our results here are not asymptotic.

We also investigate the scenario where Eve doesn't have parsing information and cannot parse the bitstream into the separate codewords. This will be the case, for example if Eve acquires only the whole bitstream, not necessarily in real--time, without the log of the network traffic. Alternatively, it acquires an encrypted file after it was saved to the disk. In this case, when we assume that the length of transmission is infinite, we show that that the best achievable rates of both the key and the transmission are given by the Huffman length of the source. In contrast to the results described in the previous paragraph, the results in this scenario are asymptotic in the sense that the probability that the system is not secure is zero when the transmission length is infinite. Note that the length of the transmission was not an issue in the mentioned previous works since block coding was used. Therefore, the block length was known a-priori to Eve and leaked no information.


In the following two sections we deal with the real--time and causal setting, receptively. Each section begins with a formal definition of the relevant problem.


\section{Real--Time Full Secrecy}\label{Sec:RealTimeSecrecy}
We begin with notation conventions. Capital letters represent scalar random variables (RV's), specific
realizations of them are denoted by the corresponding lower case letters and their alphabets -- by calligraphic letters. For $i < j$ ($i$, $j$ ñ- positive integers), $x^j_i$ will denote the vector $(x_i,\ldots, x_j)$, where for $i = 1$ the subscript will be omitted.
For two random variables $X,Y$, with alphabets $\calX,\calY$, respectively and joint probability distribution $\cbr{p(x,y)}$, the average instantaneous codeword length of $X$ conditioned on $Y=y$ will be given by

\vspace{-15pt}
\begin{align}
    L(X|Y=y) \eqd \min_{l(\cdot)\in\calA_{\calX}}\left\{\sum_{x\in\calX}P(x|y)l(x) \right\}.
\end{align}
where $\calA_{\calX}$ is the set of all possible length functions $l:\calX\to \mathbb{Z}^+$ that satisfy Kraft's inequality for alphabet of size $|\calX|$.
$L(X|Y=y)$ is obtained by designing a Huffman code for the probability distribution $P(x|y)$.
With the same abuse of notation common for entropy, we let $L(X|Y)$ denote the expectation of $L(X|Y=y)$ with respect to the randomness of $Y$. The Huffman length of $X$ is given by $L(X)$.

In this section, the following real-time source coding problem is considered: Alice, wishes to losslessly transmit the output of a DMS $X$ with probability mass function $P_X(x)$ to Bob. The communication between Alice and Bob is intercepted by Eve. Alice and Bob operate without delay. When Alice observes $X_t$ she encodes it by an instantaneous code and transmits the codeword to Bob through a clean digital channel. Bob decodes the codeword and reproduces  $X_t$. A communication stage is defined to start when the source emits $X_t$ and ends when Bob reproduces $X_t$, i.e., Bob cannot use future transmissions to calculate $X_t$. We will assume that both Alice and Bob have access to a completely random binary sequence, $\bu=(u_1,u_2,\ldots)$, which is independent of the data and will be referred to as the key. Let $m_1,m_2,\ldots,m_n$, $m_i\in\mathbbm{N}$ be a non decreasing sequence of positive integers. At stage $t$, Alice uses $l_{K_t}\eqd m_{t}-m_{t-1}$ bits that were not used so far from the key sequence. Let $K_t\eqd (u_{m_{t-1}+1},\ldots,u_{m_{t}})$ denote the stage $t$ key. The parsing of the key sequence up to stage $t$ should be the same at Alice and Bob. This can be done ``on the fly'' through the data already known to both Alice and Bob from the previous stages.  We define the key rate to be  $R_K= \limsup_{n\to\infty}\frn\sum_{t=1}^n \bE l_{K_t}$. We will also assume that Alice has access, at each stage, to a private source of randomness $\{V_t\}$, which is i.i.d and independent of the source and the key. Neither Bob nor Eve have access to $\cbr{V_t}$.

Let $\calZ$ be the set of all finite length binary strings. Denote Alice's output at stage $t$ by $Z_t\in\calZ$ and let $B_t$ denote the unparsed sequence, containing $l_{B_t}$ bits, that were transmitted so far up to the end of stage $t$. The rate of the encoder is defined by $R\eqd  \limsup_{n\to\infty}\frn\bE l_{B_n}$.

Given the keys up to stage $t$, $K^t$, Bob can parse $B_k$ into $Z_1,\ldots,Z_t$ for any $k\geq t$. The legitimate decoder is thus a sequence functions $X_t=g_t(K^t,Z^t)$.

As discussed in the Introduction, we will treat two security models. In the first model we will assume that Eve can detect when each stage starts, i.e., it can parse $B_t$ into $Z_1,\ldots,Z_t$. In the second model, we will assume that Eve intercepts the whole bitstream $B_n$ (assuming a total of $n$ stages) but has no information on actual parsing of $B_n$ into $Z_1,\ldots,Z_n$. These models are treated in the following two subsections.

\subsection{Eve Has Parsing Information}
In this subsection we assume that Eve can parse $B_n$ into $Z_1,Z_2,\ldots,Z_n$. In order for the system to be fully secure, following \cite{Shannon1949}, we will require that for any $k,m,n$, $P(X^k|Z_m^n)=P(X^k)$, i.e., acquiring any portion of the transmission leaks no information on the source, which was not known to Eve in advance.

The most general real--time encoder is a sequence of functions $Z_t = f_t(K^t,V_t,X^t)$. In this paper, we will treat only a subclass of encoders that satisfy the Markov chain
\begin{align}
	X_t\leftrightarrow Z^t\leftrightarrow K^{t-1} \label{eq:RT_Markov}.
\end{align}
Namely, given the past and current encoder outputs, the current source symbol, $X_t$, does not reduce the uncertainty regarding the past keys. We claim that this constraint, in the framework of complete security is relatively benign and, in fact,  any encoder that calculates a codeword (possibly using the whole history of the source and keys, i.e., with the most general encoder structure), say $\hZ_t$, and then outputs $Z_t=\hZ_t \oplus K_t$ will satisfy this constraint. Such a structure seems natural for one--time pad encryption. Another example of encoders that will satisfy such a constraint are encoders with the structure $Z_t = f_t(K_t,V_t,X_t,Z^{t-1})$ (we omit the proof this structure will induce the Markov chain due to space limitations). The main result of this subsection is the following theorem:
\begin{theorem} \label{Thm:RT}
 There exists a pair of fully secure real--time encoder and decoder if and only if $R_K \geq L(X)$.
\end{theorem}
This theorem is in the spirit of the result of \cite{Shannon1949}, where the entropy is replaced by the Huffman length due to the real-time constraint. As discussed in the introduction, variable--rate coding is not an option when we want the communication to be fully secure. This means that the encoder should either output constant length (short) blocks or have the transmission length independent of the source symbol in some other way. Clearly, with constant length blocks, the rate of a lossless encoder cannot be as low as $L(X)$ for all possible memoryless sources. The rate of the key, however, can be as low as $L(X)$. In the proof of the direct part of Theorem \ref{Thm:RT}, we show that a constant rate encoder with block length corresponding to the longest Huffman codeword achieves this key rate. The padding is done by random bits from the encoder's private source of randomness.  Note, however, that if both the key rate and encoder rate are $\log|X|$, lossless fully secure communication is trivially possible. Although Theorem \ref{Thm:RT} does not give a lower bound on the rate of the encoder, the above discussion suggests that there is a trade-off between the key rate and the possible encoder rate that will allow secure lossless communication. Namely, there is a set of optimal rate pairs, $(R,R_K)$, which are possible.  We prove Theorem \ref{Thm:RT} in the following subsections.

\subsubsection{Converse}

For every lossless encoder--decoder pair that satisfies the security constraint and \eqref{eq:RT_Markov}, we lower bound the key rate as follows:
\allowdisplaybreaks{\begin{align}
 	\sum_{t=1}^n \bE l_{K_t} &= \sum_{t=1}^n L(K_t) \nl
	&\geq \sum_{t=1}^n L(K_t|K^{t-1},Z^t)\label{eq:RT_Key0}\\
	&= \sum_{t=1}^n L(K_t,X_t|K^{t-1},Z^t)\label{eq:RT_Key1}\\
	&\geq \sum_{t=1}^n L(X_t|K^{t-1},Z^t)\label{eq:RT_Key2}\\
	&= \sum_{t=1}^n L(X_t|Z^t)\label{eq:RT_Key3}\\
	& =\sum _{t=1}^n L(X_t)\label{eq:RT_Key4}\\
    &=nL(X).
\end{align}}
The first equality is true since the key bits are incompressible and therefore the Huffman length is the same as the number of key bits. \eqref{eq:RT_Key0} is true since conditioning reduces the Huffman length (the simple proof of this is omitted). \eqref{eq:RT_Key1} follows since $X_t$ is a function of $(K^t,Z^t)$ (the decoder's function) and therefore, given $(K^{t-1},Z^t)$, the code for $K_t$ also reveals $X_t$. \eqref{eq:RT_Key2} is true since with the same conditioning on $(K^{t-1},Z^t)$, the instantaneous code of $(K_t,X_t)$ cannot be shorter then the instantaneous code of $X_t$. \eqref{eq:RT_Key3} is due to \eqref{eq:RT_Markov} and finally, \eqref{eq:RT_Key4} is true by the security model.
We therefore showed that $R_K\geq L(X)$.
\subsubsection{Direct}\label{Sec:RT_Direct}
We construct an encoder--decoder pair that are fully secure with $R_K = L(X)$. Let $l_{max}$ denote the longest Huffman codeword of $X$. We know that $l_{max}\leq |X|-1$. The encoder output will always be $l_{max}$ bits long and will be built from two fields. The first field will be the Huffman codeword for the observed source symbol $X_t$. Denote its length by $l(X_t)$. This codeword is then XORed with $l(X_t)$ key bits. The second field will be composed of $l_{max}-l(X_t)$ random bits (taken from the private source of randomness) that will pad the encrypted Huffman codeword to be of length $l_{max}$. Regardless of the specific source output, Eve sees constant length codewords composed of random uniform bits. Therefore no information about the source is leaked by the encoder outputs. When Bob receives such a block, it starts XORing it with key bits until it detects a valid Huffman codeword. The rest of the bits are ignored. Obviously, the key rate which is needed is $L(X)$.

\subsection{Eve Has No Parsing Information}
In this subsection, we relax our security assumptions and assume that Eve observes the whole transmission from Alice to Bob, but has no information on how to parse the bitstream $B_n$ into $Z_1,\ldots, Z_n$. Although it is not customary to limit the eavesdropper in any way in information--theoretic security, this limitation has a practical motivation, as discussed in the Introduction.

We will require that the following holds for every $t$ and every $x\in\calX$:
\begin{align}
 P(X_t=x|B_n) \xrightarrow[n\to\infty]{} P_X(X_t=x)~~ a.s.\label{eq:RT_SecCon}
\end{align}
This means that when the bitstream is long enough, the eavesdropper does not learn from it anything about the source symbols. Note that the encoder from Section \eqref{Sec:RT_Direct} trivially satisfies this constraint since it was a constant block length encoder and the bits within the block where encrypted by a one--time pad. We will see that with the relaxed secrecy requirement we can reduce the rate of the encoder to be the same as the rate of the key. In this section we deal with encoders that satisfy $X_t\leftrightarrow B_n\leftrightarrow K^{t-1}$.
The discussion that followed the constraint \eqref{eq:RT_Markov} is valid here as well. We have the following theorem:
\begin{theorem}\label{Thm:RT_NoParse}
	There exists a lossless encoder--decoder pair that satisfies the secrecy constraints \eqref{eq:RT_SecCon} if and only if	
 	$R \geq L(X), R_K \geq L(X)$.
\end{theorem}
The fact that $R\geq L(X)$ is trivial since we deal with a real time lossless encoder. However, unlike the case of Theorem \ref{Thm:RT}, here it can be achieved along with $R_K\geq L(X)$. The proof of the bound on $R_K$  follows the proof of the previous section up to \eqref{eq:RT_Key3} by replacing $Z^t$ by $B_n$. We have: $R_K\geq \frn\sum_{t=1}^n L(X_t|B_n)$.
Now, since $P(X_t|B_n)\to P(X_t)~a.s.$ we have that $L(X_t|B_n)\to L(X_t) ~a.s.$.
The direct part of the proof is achieved by separation. We first encode $X_t$ using a Huffman code and then XOR the resulting bits with a one time pad. Therefore, both the encoder and key rate of this scheme are equal to $L(X)$. We need to show that \eqref{eq:RT_SecCon} holds. We outline the idea here. The bits of $B_n$ are independent of $X_t$ since we encrypted them with a one-time pad. Let $l_{B_n}$ represent the number of bits in $B_n$. Since $B_n$ is encrypted we have $X_t\to l_{B_n}\to B_n$. Therefore, we have that $P(X_t|B_n)=P(X_t|B_n,l_{B_n})=P(X_t|l_{B_n})$. From the law of large numbers, $l_{B_n}\to nL(X)~a.s$. But if $l_{B_n}=nL(X)$ then $l_{B_n}$ leaks no information about  $X_t$ (since this $nL(X)$ is known a-priori to Eve). The full proof resembles the martingale proof of the strong law of large numbers and can be found in \cite{MeSecrecy2012}.

\noindent\textit{Discussion}:
Unlike Theorem \ref{Thm:RT}, Theorem \ref{Thm:RT_NoParse} addresses the rate of the encoder as well as the rate of the key. The result here is asymptotic since only when the bitsream is long enough we have the independence of $X_t$ from $B_n$. It can be shown that the probability that $B_n$ reveals information on $X_t$ vanishes exponentially fast with $n$. Note that if instead of defining the security constraint as in \eqref{eq:RT_SecCon}, we would have required that for every $n,t$, $P(X_t|B_n)=P(X_t)$ then a counterpart of Theorem 1 will hold here. However, the encoder will, as in the direct part of Theorem \ref{Thm:RT} proof, work in constant rate.

\section{Causal Rate Distortion with Security Constraints and SI}\label{Sec:CausalRD}
In this section, we extend the work of \cite{NeuhoffGilbert1982},\cite{TsachyNeri2005} to include secrecy constraints. We consider the following source model: Alice, Bob, and Eve observe sequences of random variables $X^n$, $Y^n$, and $W^n$ respectively which take values over discrete alphabets $\calX,\calY,\calW$, respectively.  $(X^n, Y^n, W^n)$ are distributed according to a joint distribution $p(x^n,y^n,w^n) = \prod_{t=1}^nP(x_t)P(y_t|x_t)P(w_t|y_t)$, i.e., the triplets $(X_t,Y_t,W_t)$ are created by a DMS with the structure $X\leftrightarrow Y \leftrightarrow W$. $(Y^n,W^n)$ are the SI sequences seen by Bob and Eve respectively. Unlike \cite{Prab-Ramch1997}, \cite{VillardPaint2011}, we will treat in this paper only the case of degraded SI. This model covers the scenarios where no SI is available or is available only to Bob as special cases.
Both Alice and Bob have access to a shared secret key denoted by $K$, $K\in\{0,1,2\ldots,M_k\}$ which is independent of the source.

Let $\hcalX$ be Bob's reproduction alphabet and let $d: \calX\times\hcalX\to  [0,\infty)$, $d_{min}=\min_{x,\hx}d(x,\hx)$. Finally, let $d(x^n,\hx^n) = \frn\sum_{t=1}^n d(x_t,\hx_t)$.
Alice encodes $X^n$, using the key, $K$, and creates a bit sequence $\bZ=Z_1,Z_2\ldots$ which is transmitted through a clean channel to Bob. Bob uses $(K,Y^n, \bZ)$ to create an estimate sequence, $\hX^n$, such that $\bE d(X^n,\hX^n)\leq D$.  We allow the decoder to fail and declare an error with a vanishing probability of error. Namely, for every $\delta>0$ there exists $n$ large enough such that the probability of error is less than $\delta$.

We assume that Eve intercepts the transmitted bits, $\bZ$. The security of the system is measured by the uncertainty of Eve with regard to the source sequence, measured by $\frn H(X^n|W^n,\bZ)$.
As in \cite{NeuhoffGilbert1982}, we call the cascade of encoder and decoder a reproduction coder. We say that a reproduction function is causal relative to the source if
\begin{align}
    \hX_t = f_t(X_{-\infty}^{\infty},K) = f_t(\tilde{X}_{-\infty}^{\infty},K) \text{ if } X_{-\infty}^{t}=\tilde{X}_{-\infty}^{t}
\end{align}
Note that we did not restrict the use of the key to be causal in any sense. Moreover, this definition does not rule out arbitrary delays and real--time is not considered here.  We will only treat the SI model covered in \cite{TsachyNeri2005} where $Y^n$ in not used in the reproduction of $\hX^n$ but can be used for the compression of $\hX^n$. More complicated models will be treated in \cite{MeSecrecy2012}. 
A causal reproduction coder is characterized by a family of reproduction functions $\{f_k\}_{k=1}^{\infty}$, such that the reproduction $\hX_k$ of the $k$th source output $X_k$ is given by $\hX_k=f_k(K,X^k)$. If the decoder declares an error, we will have $\hX_k\neq f_k(K,X^k)$. The probability of this event is the probability of decoder error. The average distortion of an encoder-decoder pair with an induced reproduction coder $\{f_k\}$ is defined by
\begin{align}
 	d(\cbr{f_k}) = \limsup_{n\to\infty}\bE\sbr{d(X^n,\hX^n)}.
\end{align}
The encoder's rate is defined by $R = \frn\limsup_{n\to\infty}H(\bZ).$

Let $\calR$ denote the set of positive quadruples $(R,R_K,D,h)$ such that for every $\epsilon>0,\delta>0$ and sufficiently large $n$, there exists an encoder and a decoder whose probability of error is less than $\delta$, inducing a causal reproduction coder satisfying:
\begin{align}
     \frn H(\bZ) &\leq R+\epsilon \label{eq:R}\\
    \frn H(K) &\leq R_K +\epsilon\label{eq:R_k}\\
    \frn\sum_{t=1}^n\bE\rho(X_t,\hX_t)&\leq D +\epsilon\\
    \frn H(X^n|W^n,\bZ)&\geq h -\epsilon \label{eq:Equivo}
\end{align}
Let $r_{x|y}(D)$ be the optimum performance theoretically attainable function (OPTA) from \cite{TsachyNeri2005} for the case where the SI is available only at the decoder. Namely,
\begin{align}
	 r_{x|y}(D) = \min_{f: \boldmath{E} d(X,f(X))\leq D}H(f(X)|Y)	
\end{align}
and let $\overline{r_{x|y}}(\cdot)$ denote the lower convex envelope of $r_{x|y}(\cdot)$.
We have the following theorem.

\begin{theorem}
    $(R,R_k,D,h)\in\calR$ if and only if
    \begin{align}
        h&\leq H(X|W), D \geq D_{min}, R\geq\overline{r_{x|y}}(D),\nt\\
        R_k&\geq h - H(X|W) +  \overline{r_{x|y}}(D). \label{eq:Theorem}
    \end{align}
    If $h - H(X|W) +  \overline{r_{x|y}}(D)\leq 0$, no encryption is needed.
\end{theorem}
It is seen from the theorem, that separation holds in this case. The direct part of this proof is therefore straightforward: First, quantize the source within distortion $D$ by the scheme given in \cite{TsachyNeri2005}. As was shown in \cite{TsachyNeri2005}, this step requires time-sharing no more than two memoryless quantizers. Now use Slepian--Wolf encoding to encode the resulting quantized symbols given the SI at Bob. Finally, use a one--time pad of $n(h - H(X|W) +  \overline{r_{x|y}}(D))$ bits on the block describing the bin number.

We now proceed to prove the converse part, starting with lower bounding the encoding rate. For any $k$, let  $\tX_k=f_k(K,X^k)$. $\tX_k$ are equal to $\hX_k$ when there is no decoding error. Since the probability of decoder failure vanishes, we have from Fano's inequality (\cite{cover}) that for every $\epsilon>0$, there exists $n$ large enough such that $H(\tX^n|K,Y^n,\bZ)\leq n\epsilon$.

For $n$ large enough and every encoder and decoder pair that induce a causal reproduction coder and satisfy \eqref{eq:R}-\eqref{eq:Equivo} the following chain of inequalities hold:
\begin{align}
	nR &\geq H(\bZ) \nl
	&\geq H(\bZ|K,Y^n) - H(\bZ|K,\tX^n,Y^n)\nl
	&=I(\tX^n;\bZ|K,Y^n)\nl
	&=H(\tX^n|K,Y^n)-H(\tX^n|K,Y^n,\bZ)\nl
	&\geq H(\tX^n|K,Y^n)-n\epsilon\label{eq:SIConverse2}\\
	&=\sum_{t=1}^n H(\tX_t|K,\tX^{t-1},Y^n) -n\epsilon\nl
	&\geq \sum_{t=1}^n H(\tX_t|K,\tX^{t-1},X^{t-1},Y^n) -n\epsilon\nl
	&=\sum_{t=1}^n H(\tX_t|K,X^{t-1},Y^n)-n\epsilon\nl
	&=\sum_{t=1}^n H(f_t(X^{t-1},K,X_t)|K,X^{t-1},Y^n)-n\epsilon
\end{align}
where \eqref{eq:SIConverse2} follows from Fano's inequality. From here, using the independent of the key and the source and following the steps used in \cite[Appendix, eq. A.11]{TsachyNeri2005} we can show that $R\geq \overline{r_{x|y}}(D)$.
The key rate can be lower bounded as follows:
\begin{align}
&nR_K = H(K)\nl
&\geq H(K|\bZ,W^n)\nl
&=I(X^n;K|W^n,\bZ) + H(K|X^n,W^n,\bZ)\nl
&=H(X^n|W^n,\bZ)-H(X^n|K,W^n,\bZ)+H(K|X^n,W^n,\bZ)\nl
&\geq nh-H(X^n|K,W^n,\bZ)+H(K|X^n,W^n,\bZ)\nl
&\geq nh-H(X^n|K,W^n,\bZ)\label{eq:SIConverseStart}
\end{align}
We continue by focusing on $H(X^n|K,W^n,\bZ)$:
\begin{align}
&H(X^n|K,W^n,\bZ) \nl
&= I(X^n,Y^n|K,W^n,\bZ)+H(X^n|K,Y^n,W^n,\bZ)\nl
&\leq H(Y^n|W^n)-H(Y^n|K,X^n,W^n,\bZ)+H(X^n|K,Y^n,\bZ)\label{eq:SIConverse04}\\
&= H(Y^n|W^n)-H(Y^n|X^n,W^n)+H(X^n|K,Y^n,\bZ)\label{eq:SIConverse4}\\
&= I(X^n;Y^n|W^n)+H(X^n|K,Y^n,\tX^n,\bZ) \nl
&~~~~~~~+ I(X^n;\tX^n|K,Y^n,\bZ)\nl
&\leq nI(X;Y|W)+H(X^n|K,Y^n,\tX^n,\bZ) +n\epsilon \label{eq:SIConverse7}\\
&\leq n(H(X|W)-H(X|Y)) +H(X^n|K,Y^n,\tX^n) \label{eq:SIConverse6}
\end{align}
where in \eqref{eq:SIConverse04} we used the degraded structure of the source. \eqref{eq:SIConverse4} is true since $\bZ$ is a function of $(K,X^n)$ and $K$ is independent of the souce. \eqref{eq:SIConverse7} is true by Fano's inequality and the fact that $\tX^n$ is a function of $(K,X^n)$. Focusing on the last term of \eqref{eq:SIConverse6} we have
\begin{align}
&H(X^n|K,Y^n,\tX^n) = H(X^n|K,Y^n) -I(X^n;\tX^n|K,Y^n)\nl
&= nH(X|Y)-I(X^n;\tX^n|K,Y^n)\nl
&= nH(X|Y)-H(\tX^n|K,Y^n)+H(\tX^n|K,X^n,Y^n)\nl
&\leq nH(X|Y)-H(\tX^n|K,Y^n) +n\epsilon \label{eq:SIConverse5}\\
&\leq nH(X|Y)-\overline{r_{x|y}}(D)+\epsilon \label{eq:SIConverseEnd}
\end{align}
where \eqref{eq:SIConverse5} is true since $\hX^n$ is a function of $K,X^n$ through the reproduction coders and Fano's inequality. Finally the last line follows from \eqref{eq:SIConverse2}.
Combining \eqref{eq:SIConverseEnd} with \eqref{eq:SIConverse6} into \eqref{eq:SIConverseStart} we showed that $R_K \geq h - H(X|W) +  \overline{r_{x|y}}(D)$.
\bibliographystyle{IEEEtran}
\bibliography{PhDBib-1}

\end{document}